\documentstyle[aps]{revtex}

\begin{document}
\title{New criteria of separability for tripartite and more high dimensional
multipartite qubit density matrixes }
\author{Zai-Zhe Zhong}
\address{Department of Physics, Liaoning Normal University, Dalian 116029, Liaoning,\\
China. E-mail: zhongzaizheh@hotmail.com}
\maketitle

\begin{abstract}
In this Letter we find the new criteria of separability of multipartite
qubit density matrixes. Especially, we discuss in detail the criteria of
separability for tripartite qubit density matrixes. We find the sufficient
and necessary conditions of separability for tripartite qubit density
matrixes, and give two corollaries. The second corollary can be taken as the
criterion of existence of entanglement for tripartite qubit density
matrixes. In concrete application, its steps are quite simple and easy to
operate. Some examples, discussions and the generalization to more high
dimensional multipartite qubit density matrixes are given.

PACC numbers: 03.65.Ud, 03.65.Bz, 03.67.-a, 03.67.Hk.
\end{abstract}

\smallskip It is known that in modern quantum mechanics, especially in the
quantum information theory, to find the criteria of separability of density
matrixes is an important task in recent years. The first important result is
the well-known positive partial transposition(PPT, Peres-Horodecki)
criteria[1,2$]$ for $2\times 2$ and $2\times 3$ systems. Recently there are
many studies concerned one after other, e.g. see [3-7$],$ about the criteria
of separability for multipartite systems see [8-15$].$ As the simplest case
of multipartite systems, the criteria of separability of tripartite qubit
density matrixes are yet studied, e.g. see [$10,12$], and the classification
problem of tripartite qubit pure-states has been completed[16$]$. However,
for multipartite qubit systems, especially for tripartite qubit
mixed-states, we always hope to find the criteria of separability which are
more simple and easy to operate. Since there always are many known results
about the separability for bipartite qubit density matrixes, then we should
consider such a problem: Can the criteria of separability of multipartite
qubit density matrixes be reduced, in some cases, to the criteria for
several bipartite qubit density matrixes? If we can accomplish this point,
then we can ascertain the separability for multipartite qubit density
matrixes with the aid of various known results about bipartite qubit density
matrixes. In this Letter we give such criteria.

The tripartite qubit states are most simple and important multipartite
states for application, hence in the first place, we study in detail the
special criteria of separability for tripartite qubit density matrixes. It
is shown that the necessary and sufficient condition of separability of a
tripartite qubit density matrix is that there is a appropriate decomposition
and several special reduced bipartite qubit density matrixes must be
separable. Two corollaries are given, and the second corollary, in fact, can
be taken as the criterion of existence of entanglement for a tripartite
qubit density matrix. In concrete application, its steps are quite simple
and easy to operate. Some examples, the discussion about conditions in
theorem and corollaries, and the generalization to more high dimensional
multipartite qubit density matrixes are given.

In this Letter, we shall repeatedly use the definition of the separability
of a (bipartite or multipartite) state $\rho ,$ i.e. $\rho $ is separable if
and only if there is a decomposition that $\rho =\sum\limits_\alpha p_\alpha
\rho _\alpha ,0\leq p_\alpha \leq 1,\sum\limits_\alpha p_\alpha =1$ such
that $\rho _\alpha $ is a separable pure-state for any $\alpha $.

Let $\mid i_A>,\;\mid j_B$ 
\mbox{$>$}%
and $\mid k_C>(i,j,k=0,1)$ be the natural bases, they span the Hilbert
spaces $H_A,\;H_B$ and $H_C$, respectively. A tripartite qubit density
matrix $\rho $ as an operator acting upon $H_A\otimes H_B\otimes H_C$ can be
written as 
\begin{equation}
\rho =\sum_{i,j,k,r,s,t=0}^1\left[ \rho \right] _{ijk,rst}\mid i_A>\mid
j_B>\mid k_C><r_A\mid <s_B\mid <t_C\mid
\end{equation}
where $\left[ \rho \right] _{ijk,rst}$ are the entries of matrix $\rho .$
Now we define six 4$\times 4$ matrixes $\rho _{\left( A,B\right) },\rho
_{\left( A,C\right) },\rho _{\left( B,C\right) },\rho _{\left( A,BC\right)
},\rho _{\left( B,CA\right) }$ and $\rho _{\left( C,AB\right) }$ as follows.
First, 
\begin{equation}
\rho _{\left( A,B\right) }=tr_C\left( \rho \right) ,\;\rho _{\left(
A,C\right) }=tr_B\left( \rho \right) ,\;\rho _{\left( B,C\right)
}=tr_A\left( \rho \right)
\end{equation}
Next, $\rho _{\left( A,BC\right) },\rho _{\left( B,CA\right) }$ and $\rho
_{\left( C,AB\right) }$, respectively, are defined by 
\begin{eqnarray}
\left[ \rho _{\left( A,BC\right) }\right] _{ij,rs} &=&\left[ \rho \right]
_{ijj,rss}+\left[ \rho \right] _{ij\left( 1-j\right) ,rs\left( 1-s\right)
},\;\left[ \rho _{\left( B,CA\right) }\right] _{ij,rs}=\left[ \rho \right]
_{jij,srs}+\left[ \rho \right] _{\left( 1-j\right) ij,\left( 1-s\right) rs} 
\nonumber \\
\left[ \rho _{\left( C,AB\right) }\right] _{ij,rs} &=&\left[ \rho \right]
_{jji,ssr}+\left[ \rho \right] _{j\left( 1-j\right) i,s\left( 1-s\right) r}
\end{eqnarray}

{\bf Lemma. }$\rho _{\left( A,B\right) },\rho _{\left( A,C\right) },\rho
_{\left( B,C\right) },\rho _{\left( A,BC\right) },\rho _{\left( B,CA\right)
} $ and $\rho _{\left( C,AB\right) }$ all are bipartite density matrixes.

{\bf Proof. }Since{\bf \ }$\rho _{\left( A,B\right) },\rho _{\left(
A,C\right) },\rho _{\left( B,C\right) }$ all are ordinary reductions of $%
\rho ,$ then lemma holds, we only need to make the proofs for $\rho _{\left(
A,BC\right) },\rho _{\left( B,CA\right) }$ and $\rho _{\left( C,AB\right) }$.

In the first place, we assume that $\rho $ is a pure-state, $\rho =\mid \Psi
><\Psi \mid ,$ where $\Psi =\sum\limits_{i,j,k=0}^1c_{ijk}\mid i_A>\mid
j_B>\mid k_C>$ $\in H_A\otimes H_B\otimes H_C$ is normalized, i.e. 
\begin{equation}
\sum\limits_{i,j,k=0,1}\left| c_{ijk}\right| ^2=1
\end{equation}
We take two form bases $\mid m_X>$ and $\mid n_Y>(m,n=0,1)$ and define 
\begin{equation}
\Phi _{(A,BC)}=\left( \eta _{\left( A,BC\right) }\right)
^{-1}\sum_{m,n=0,1}c_{mnn}\mid m_X>\mid n_Y>,\;\Phi _{(A,B\stackrel{\vee }{C}%
)}=\left( \eta _{(A,B\stackrel{\vee }{C})}\right)
^{-1}\sum_{m,n=0,1}c_{mn\left( 1-n\right) }\mid m_X>\mid n_Y>
\end{equation}
where $\eta _{\left( A,BC\right) }=\sqrt{\sum\limits_{m,n=0,1}\left|
c_{mnn}\right| ^2},\;\eta _{(A,B\stackrel{\vee }{C})}=\sqrt{%
\sum\limits_{m,n=0,1}\left| c_{mn\left( 1-n\right) }\right| ^2}$, then $\Phi
_{(A,BC)}$ and $\Phi _{(A,B\stackrel{\vee }{C})}$ both are normalized
bipartite qubit pure states. From Eqs.(4) and (5) we have 
\begin{eqnarray}
\rho _{\left( A,BC\right) } &=&\eta _{\left( A/BC\right) }^2\rho
_{(A/BC)}+\;\eta _{(A/B\stackrel{\vee }{C})}^2\rho _{(A/B\stackrel{\vee }{C}%
)},\;\rho _{(A/BC)}=\mid \Phi _{(A/BC)}><\Phi _{(A/BC)}  \nonumber \\
\;\rho _{(A/B\stackrel{\vee }{C})} &=&\mid \Phi _{(A/B\stackrel{\vee }{C}%
)}><\Phi _{(A/B\stackrel{\vee }{C})}\mid ,\;\eta _{\left( A,BC\right)
}^2+\eta _{(A,B\stackrel{\vee }{C})}^2=1
\end{eqnarray}
since $\rho _{(A/BC)}$ and $\rho _{(A/B\stackrel{\vee }{C})}$ both are
bipartite pure-states, $\rho _{\left( A,BC\right) }$ is a bipartite density
matrix (a mixed-state).

Secondly, if $\rho =\sum\limits_\alpha p_\alpha \rho _\alpha $ is a
mixed-state, where the real numbers 0$\leq p_\alpha \leq 1$ satisfy $%
\sum\limits_\alpha p_\alpha =1,$ and every $\rho _\alpha $ is a pure-state,
then we can obtain $\eta _{\alpha \left( A,BC\right) },\rho _{\alpha
(A/BC)},\eta _{\alpha (A,B\stackrel{\vee }{C})}$ and $\rho _{\alpha (A/B%
\stackrel{\vee }{C})}$ for every $\rho _\alpha $ . From Eqs.(5) and (6), we
have 
\begin{equation}
\rho _{\left( A,BC\right) }=\sum_\alpha \left( \zeta _{\alpha (A/BC)}\rho
_{\alpha (A/BC)}+\zeta _{\alpha (A/B\stackrel{\vee }{C})}\rho _{\alpha (A/B%
\stackrel{\vee }{C})}\right)
\end{equation}
where $\zeta _{\alpha (A/BC)}=p_\alpha \eta _{\left( A/BC\right) }^2,\zeta
_{\alpha (A/B\stackrel{\vee }{C})}=p_\alpha \eta _{(A/B\stackrel{\vee }{C}%
)}^2,\sum_\alpha \left( \zeta _{\alpha (A/BC)}+\zeta _{\alpha (A/B\stackrel{%
\vee }{C})}\right) =1.$ Since all $\rho _{\alpha (A/BC)},\rho _{\alpha (A/B%
\stackrel{\vee }{C})}$ are bipartite qubit density matrixes, $\rho _{\left(
A,BC\right) }$ is a bipartite qubit density matrix (a mixed state).
Similarly, we can prove the cases for $\rho _{\left( B,CA\right) }$ and $%
\rho _{\left( C,AB\right) }.$ {\bf QED}

From the above lemma we know that $\rho _{\left( A,BC\right) },\rho _{\left(
B,CA\right) }$ and $\rho _{\left( C,AB\right) }$, in fact, are yet some
special reductions of $\rho .$

{\bf Theorem.} The necessary and sufficient condition of separability for a
tripartite qubit density matrix $\rho $ is that there is a decomposition $%
\rho =\sum\limits_\alpha p_\alpha \rho _\alpha ,0\leq p_\alpha \leq 1,$ $%
\sum\limits_\alpha p_\alpha =1,$ such that the bipartite qubit density
matrixes $\left( \rho _\alpha \right) _{\left( A,B\right) },\left( \rho
_\alpha \right) _{\left( A,C\right) },\left( \rho _\alpha \right) _{\left(
B,C\right) },\left( \rho _\alpha \right) _{\left( A,BC\right) },$ $\left(
\rho _\alpha \right) _{\left( B,CA\right) }$ and $\left( \rho _\alpha
\right) _{\left( C,AB\right) }$ all are separable.

{\bf Proof. }In the first place, we prove that when $\rho $ is a pure-state $%
\rho =\mid \Psi ><\Psi \mid ,$ where $\Psi
=\sum\limits_{i,j,k=0,1}c_{ijk}\mid i_A>\mid j_B>\mid k_C>\in H_A\otimes
H_B\otimes H_C,$ then this theorem holds (in this case the decomposition is
just the identity $\rho =\rho )$. In fact, in this case the separability of $%
\rho $ is just separabolity of $\Psi $. It easily directly verified that $%
\rho $ is A-BC separable (i.e. $\Psi =\Psi _A\Psi _{BC},$ $\Psi _A\in
H_A,\Psi _{BC}\in H_B\otimes H_C),$ if and only if the rank of the following
matrix 
\begin{equation}
\begin{array}{c}
\left( i_A=0\rightarrow \right) \\ 
\left( i_A=1\rightarrow \right)
\end{array}
\left[ 
\begin{array}{cccc}
c_{000} & c_{001} & c_{010} & c_{011} \\ 
c_{100} & c_{101} & c_{110} & c_{111}
\end{array}
\right]
\end{equation}
is less than 2, i.e. in which the determinant of any one submatrix must
vanish. Next, we notice that since 
\begin{equation}
\rho _{\left( A,B\right) }=\mid \Psi _{A/B0}><\Psi _{A/B0}\mid +\mid \Psi
_{A/B1}><\Psi _{A/B1}\mid ,\rho _{\left( A,C\right) }=\mid \Psi
_{A/0C}><\Psi _{A/0C}\mid +\mid \Psi _{A/1C}><\Psi _{A/1C}\mid
\end{equation}
where $\Psi _{A/Bm}=\sum\limits_{i,j=0,1}c_{ijm}\mid i_X>\mid j_Y>,\left(
m=0,1\right) ,\Psi _{A/mC}=\sum\limits_{i,j=0,1}c_{imj}\mid i_X>\mid
j_Y>,\left( m=0,1\right) .$ By the normalization, we can write 
\begin{equation}
\rho _{\left( A,B\right) }=\eta _{A/B0}^2\Phi _{A/B0}+\eta _{A/B1}^2\Phi
_{A/B1},\rho _{\left( A,C\right) }=\eta _{A/0C}^2\Phi _{A/0C}+\eta
_{A/1C}^2\Phi _{A/1C}
\end{equation}
where the normalization factors $\eta _{A/Bm}=\sqrt{\sum\limits_{i,j=0,1}%
\left| c_{ijm}\right| ^2},\eta _{A/mC}=\sqrt{\sum\limits_{i,j=0,1}\left|
c_{imj}\right| ^2},\Phi _{A/Bm}=\left( \eta _{A/Bm}\right) ^{-1}\Psi
_{A/Bm}, $ $\Phi _{A/mC}=\left( \eta _{A/mC}\right) ^{-1}\Psi _{A/mC},$ $%
\left( m=0,1\right) .$ As for $\rho _{\left( A,BC\right) },$ it has been
defined as in Eqs.(3) and (5). We know that a bipartite state $\Phi
=\sum\limits_{i,j=0,1}d_{ij}\mid i_X>\mid j_Y>$ is separable if and only if
the determinant $\left| 
\begin{array}{cc}
d_{00} & d_{01} \\ 
d_{10} & d_{11}
\end{array}
\right| =0.$ Therefore $\rho $ is A-BC separable if and only if $,\Phi
_{A/Bm},\Phi _{A/mC},\Phi _{A/BC}$ and $\Phi _{A/B\stackrel{\vee }{C}}$ all
are separable. Since $\eta _{A/B0}^2+\eta _{A/B1}^2$ $=\eta _{A/0C}^2+\eta
_{A/1C}^2=\eta _{A/BC}^2+\eta _{A/B\stackrel{\vee }{C}}^2=1,$ this means
that $\rho $ is A-BC separable if and only if $\rho _{\left( A,B\right) },$ $%
\rho _{\left( A,C\right) }$ and $\rho _{\left( A,BC\right) }$ all are
separable. Similarly,consider matrixes 
\begin{equation}
\begin{array}{c}
\left( i_B=0\rightarrow \right) \\ 
\left( i_B=1\rightarrow \right)
\end{array}
\left[ 
\begin{array}{cccc}
c_{000} & c_{001} & c_{100} & c_{101} \\ 
c_{010} & c_{011} & c_{110} & c_{111}
\end{array}
\right] , 
\begin{array}{c}
\left( i_C=0\rightarrow \right) \\ 
\left( i_C=1\rightarrow \right)
\end{array}
\left[ 
\begin{array}{cccc}
c_{000} & c_{010} & c_{100} & c_{110} \\ 
c_{001} & c_{011} & c_{101} & c_{111}
\end{array}
\right]
\end{equation}
we know that $\rho $ is B-CA(C-AB) separable if and only if $\rho _{\left(
A,B\right) },\rho _{\left( B,C\right) },\rho _{\left( B,CA\right) }(\rho
_{\left( A,C\right) },\rho _{\left( B,C\right) },$ $\rho _{\left(
C,AB\right) }$ ) all are separable. Since $\rho $ is separable if and only
if it is A-BC,B-CA and C-AB separable simultaneously, i.e. if and only if $%
\rho _{\left( A,B\right) },\rho _{\left( A,C\right) },\rho _{\left(
B,C\right) },\rho _{\left( A,BC\right) },\rho _{\left( B,CA\right) }$ and $%
\rho _{\left( C,AB\right) }$ all are separable. This means that theorem
holds for pure-states.

Mention the point in passing, when $\rho $ is a separable pure-state, say $%
\rho =\mid \Psi _A><\Psi _A\mid \otimes \mid \Psi _B><\Psi _B\mid \otimes
\mid \Psi _C><\Psi _C\mid ,$ where $\Psi _A\equiv a_0\mid 0_A>+a_1\mid 1_A>,$
$\Psi _B\equiv b_0\mid 0_B>+b_1\mid 1_B>,$ $\Psi _C\equiv c_0\mid
0_C>+c_1\mid 1_C>,$ and $\left| a_0\right| ^2+\left| a_1\right| ^2=\left|
b_0\right| ^2+\left| b_1\right| ^2=\left| c_0\right| ^2+\left| c_1\right|
^2=1,$ then $\rho _{\left( A,B\right) }=\rho _A\otimes \rho _B,\rho _{\left(
B,C\right) }=\rho _B\otimes \rho _C,\rho _{\left( A,C\right) }=\rho
_A\otimes \rho _C,$and we have

\begin{eqnarray}
\rho _{\left( A,BC\right) } &=&\rho _A\otimes \omega _{BC},\;\omega
_{BC}=\left[ 
\begin{array}{cc}
\left| b_0\right| ^2 & \gamma _Cb_0b_1^{*} \\ 
\gamma _Cb_0^{*}b_1 & \left| b_1\right| ^2
\end{array}
\right] ,\;\gamma _C=2%
\mathop{\rm Re}%
\left( c_0c_1^{*}\right)  \nonumber \\
\rho _{\left( B,CA\right) } &=&\rho _B\otimes \omega _{CA},\;\omega
_{CA}=\left[ 
\begin{array}{cc}
\left| c_0\right| ^2 & \gamma _Ac_0c_1^{*} \\ 
\gamma _Ac_0^{*}c_1 & \left| c_1\right| ^2
\end{array}
\right] ,\;\gamma _A=2%
\mathop{\rm Re}%
\left( a_0a_1^{*}\right) \\
\;\rho _{\left( C,AB\right) } &=&\rho _C\otimes \omega _{AB},\;\omega
_{AB}=\left[ 
\begin{array}{cc}
\left| a_0\right| ^2 & \gamma _Ba_0a_1^{*} \\ 
\gamma _Ba_0^{*}a_1 & \left| a_1\right| ^2
\end{array}
\right] ,\;\gamma _B=2%
\mathop{\rm Re}%
\left( b_0b_1^{*}\right)  \nonumber
\end{eqnarray}
where $%
\mathop{\rm Re}%
$ is the real part of a complex number, and it can be verified that $\omega
_{BC},\omega _{CA}$ and $\omega _{AB}$ all are bipartite qubit density
matrixes. These results also show that our conclusion is true.

Now, we consider the case of mixed states.

{\bf Necessity.} If $\rho $ is separable or mixed-state, according to the
definition of a separable mixed-state, then there must be a decomposition $%
\rho =\sum\limits_\alpha p_\alpha \rho _\alpha ,0\leq p_\alpha \leq
1,\sum\limits_\alpha p_\alpha =1,$ and every $\rho _\alpha ${\bf \ }is
separable pure-state. Then from the above discussions, all $\left( \rho
_\alpha \right) _{\left( \bullet ,\bullet \right) },\left( \rho _\alpha
\right) _{\left( \bullet ,\bullet \bullet \right) }$ are separable
pure-states.

{\bf Sufficiency}. If there is a decomposition $\rho =\sum\limits_\alpha
p_\alpha \rho _\alpha $ and all $\left( \rho _\alpha \right) _{\left(
A,B\right) },\left( \rho _\alpha \right) _{\left( A,C\right) },$ $\left(
\rho _\alpha \right) _{\left( B,C\right) },\left( \rho _\alpha \right)
_{\left( A,BC\right) },\left( \rho _\alpha \right) _{\left( B,CA\right)
},\left( \rho _\alpha \right) _{\left( C,AB\right) }$ are separable, then
all $\rho _\alpha $ are separable, and $\rho =\sum\limits_\alpha p_\alpha
\rho _\alpha $ is separable. {\bf \ QED}

Notice that we always have the following equations, 
\begin{equation}
\rho _{\left( \bullet ,\bullet \right) }=\sum_\alpha p_\alpha \left( \rho
_\alpha \right) _{\left( \bullet ,\bullet \right) },\;\rho _{\left( \bullet
,\bullet \bullet \right) }=\sum_\alpha p_\alpha \left( \rho _\alpha \right)
_{\left( \bullet ,\bullet \bullet \right) }
\end{equation}
where $\left( \bullet ,\bullet \right) =\left( A,B\right) ,\left( A,C\right)
,\left( B,C\right) $ and $\left( \bullet ,\bullet \bullet \right) =\left(
A,BC\right) ,\left( B,CA\right) ,\left( C,AB\right) .$ From Eq.($13$) and
the theorem, obviously we have the following two corollaries.

{\bf Corollary 1.} A necessary condition of separability for a tripartite
qubit density matrix $\rho $ is that six bipartite qubit density matrixes $%
\rho _{\left( A,B\right) },\rho _{\left( A,C\right) ,}$ $\rho _{\left(
C,A\right) ,}\rho _{\left( A,BC\right) ,}\rho _{\left( B,CA\right) \text{ }}$
and $\rho _{\left( C,AB\right) }$ all are separable.

The meaning of this corollary is that if a tripartite qubit system is
separable, then any two parts in the system can completely be split, however
its inverse proposition is not true ( see {\bf Discussion }below).

{\bf Corollary 2 (Criterion of entanglement).} It is an entanglement witness
of the tripartite qubit density matrix $\rho $ that any one of six bipartite
qubit density matrixes $\rho _{\left( A,B\right) },\rho _{\left( A,C\right)
},\rho _{\left( B,C\right) },\rho _{\left( A,BC\right) },$ $\rho _{\left(
B,CA\right) }$ and $\rho _{\left( C,AB\right) }$ is entangled.

This corollary gives us a practical criterion for existence of entanglement
of $\rho $, its steps are quite simple and easy to operate. As for whether
one of six density matrix is or not entangled, which may be ascertained by
any known way, say, by the PPT(Peres-Horodecki) criteria[$1,2].$ Therefore,
say, if we find any one partial transposition of $\rho _{\left( A,B\right)
},\rho _{\left( A,C\right) },\rho _{\left( B,C\right) },\rho _{\left(
A,BC\right) },\rho _{\left( B,CA\right) }$ and $\rho _{\left( C,AB\right) }$
has a negative eigenvalue, then $\rho $ must be entangled.

{\bf Example 1.} As a simple example of pure-state, we see the GHZ-state $%
\rho =\mid \phi ><\phi \mid ,\phi =\frac 1{\sqrt{2}}\left( \mid 0_A>\mid
0_B>\mid 0_C>+\mid 1_A>\mid 1_B>\mid 1_C>\right) .$ In this case $\rho
_{\left( A,B\right) }=\rho _{\left( A,C\right) }=\rho _{\left( B,C\right) }=%
\frac 12\mid 0_X><0_X\mid \otimes \mid 0_Y><0_Y\mid +\frac 12\mid
1_X><1_X\mid \otimes \mid 1_Y><1_Y\mid ,$ they all are separable. However $%
\rho _{\left( A,BC\right) }=\rho _{\left( B,CA\right) }=\rho _{\left(
C,AB\right) }$ $=\frac 12\left( \mid 0_X>\mid 0_Y>+\mid 1_X>\mid 1_Y>\right)
\left( <0_X\mid <0_Y\mid +<1_X\mid <1_Y\mid \right) $, they all are
entangled Bell's state, this shows that the GHZ-state is entangled.

{\bf Example 2. }If $\rho $ is a $8\times 8$ matrix whose entries are
defined by 
\begin{equation}
\left[ \rho \right] _{ijk,rst}=xR_{ijk,rst}+\frac 18\left( 1-x\right) \delta
_{ir}\delta _{js}\delta _{kt}
\end{equation}
where the real variable 0$\leq x\leq 1,$ the nonvanishing entries of density
matrix $R$ are $\left[ R\right] _{010,010}=\left[ R\right] _{011,011}=\left[
R\right] _{100,100}=\left[ R\right] _{101,101}=\frac 14,\;\left[ R\right]
_{010,101}=\left[ R\right] _{011,100}=\left[ R\right] _{100,011}=\left[
R\right] _{100,101}=-\frac 14$ . It is easily verified that $\rho $ is a
tripartite qubit density matrix. Therefore 
\begin{equation}
\left[ \rho _{(A,BC)}\right] _{ij,rs}=xS_{ij,rs}+\frac 14\left( 1-x\right)
\delta _{ir}\delta _{js}
\end{equation}
where the nonvanishing entries of density matrix $S$ are $\left[ S\right]
_{01,01}=\left[ S\right] _{10,10}=-\left[ S\right] _{01,10}=-\left[ S\right]
_{10,01}=\frac 12,$ i.e. $\rho _{(A,BC)}$ just is the Werner state[$1,17].$
From [1$]$ we know that the partial transposition of $\rho _{(A,BC)}$ has
three equal eigenvalues $\frac 14\left( 1+x\right) ,$ the fourth eigenvalue
is $\frac 14\left( 1-3x\right) .$ Therefore when $x>\frac 13,$ $\rho
_{(A,BC)}$ must be entangled, this leads that $\rho $ must be entangled. If
we don't use the above way, this is not easily seen.

Generally, if a bipartite qubit entangled density matrix $R$ is given, we
can define the nonvanishing entries of $\rho $ by any one of the following
six ways: 
\begin{eqnarray}
(1)\;\left[ \rho \right] _{ijj,rss} &=&\left[ \rho \right] _{ij\left(
1-j\right) ,rs\left( 1-s\right) }=\frac 12\left[ R\right] _{ij,rs}.\;\left( 
\text{2}\right) \;\left[ \rho \right] _{jij}=\left[ \rho \right] _{\left(
1-j\right) ij,\left( 1-s\right) rs}=\frac 12\left[ R\right] _{ij,rs} 
\nonumber \\
\left( \text{3}\right) \;\left[ \rho \right] _{jji,ssr} &=&\left[ \rho
\right] _{j\left( 1-j\right) i,s\left( 1-s\right) r}=\frac 12\left[ R\right]
_{ij,rs}.\;\left( \text{4}\right) \;\left[ \rho \right] _{ij0,rs0}=\left[
\rho \right] _{ij1,rs1}=\frac 12\left[ R\right] _{ij,rs} \\
\left( \text{5}\right) \;\left[ \rho \right] _{i0j,r0s} &=&\left[ \rho
\right] _{i1j,r1s}=\frac 12\left[ R\right] _{ij,rs}.\;\left( \text{6}\right)
\;\left[ \rho \right] _{0ij,0rs}=\left[ \rho \right] _{1ij,1rs}=\frac 12%
\left[ R\right] _{ij,rs}  \nonumber
\end{eqnarray}
then the tripartite qubit density matrix $\rho $ must be entangled$.$

{\bf Example 3. }In [$18]$, a state considered (we only discuss the case of
tripartite qubit state) is as 
\begin{equation}
\rho =\sum_{rs=AB,BC,AC}p_{rs}\mid \Psi _{rs}><\Psi _{rs}\mid ,\;\mid \Psi
_{rs}>=\frac 1{\sqrt{2}}\left( \mid 0_r>\mid 1_s>+\mid 1_r>\mid 0_s>\right)
\otimes \mid 0_{rest}>
\end{equation}
where $0\leq p_{rs}\leq 1,\;\sum\limits_{rs=AB,AC,BC}p_{rs}=1.$ The
nonvanishing entries of $\rho _{\left( r,s\right) }$ are 
\begin{equation}
\left[ \rho _{\left( r,s\right) }\right] _{00,00}=\alpha _{rs},\;\left[ \rho
_{\left( r,s\right) }\right] _{11,11}=\beta _{rs},\;\left[ \rho _{\left(
r,s\right) }\right] _{22,22}=\gamma _{rs},\;\left[ \rho _{\left( r,s\right)
}\right] _{01,10}=\left[ \rho _{\left( r,s\right) }\right] _{10,01}=\frac{%
p_{rs}}2
\end{equation}
where $\left( r,s\right) =\left( A,B\right) ,\left( A,C\right) ,\left(
B,C\right) $ and the real numbers $\alpha _{rs}+\beta _{rs}+\gamma _{rs}=1$.
In order to clarify that whether $\rho $ is or not entangled, there is no
need to calculate the so-called `concurrences'[19] of $\rho _{\left(
r,s\right) }$ as did as in [18$],$ we can use the above criteria. In fact,
the partial transposition of matrix $\rho _{\left( r,s\right) }$ has four
eigenvalues $\lambda _{\left( 1\right) rs}=\beta _{rs},\lambda _{\left(
2\right) rs}=\gamma _{rs},\lambda _{\left( \pm \right) rs}=\frac 12\left(
-\alpha _{rs}\pm \sqrt{\alpha _{rs}^2+p_{rs}^2}\right) .$ It is impossible
that $p_{AB},p_{AC},p_{BC}$ all vanish simultaneously, this means that there
must be at least one of three $\lambda _{\left( -\right) rs}(rs=AB,AC,BC)$
which is negative, therefore $\rho $ must be entangled. In view of physical
point, it is true of course, since the state has contained some
`entanglement molecules'[18$].$

{\bf Discussion.} (1) It is a pity that in the Corollary 1, $\rho _{\left(
A,B\right) },\rho _{\left( A,C\right) },\rho _{\left( B,C\right) },\rho
_{\left( A,BC\right) },\rho _{\left( B,CA\right) }$ and $\rho _{\left(
C,AB\right) }$ all to be separable only is a necessary condition for
separability of $\rho .$ In fact, for some $\rho $ the six reduced matrixes
all are separable, however $\rho $ still are entangled. For instance, a
state given in [$8$] is as 
\begin{equation}
\rho =\frac 14\left( I-\sum_{i=1}^4\mid \psi _i><\psi _i\mid \right)
\end{equation}
where $\mid \psi _1>=\frac 1{\sqrt{2}}\mid 0_A>\mid 1_B>\mid \left( \mid
0_C>+\mid 1_C>\right) ,\;\mid \psi _2>=\frac 1{\sqrt{2}}\mid 1_A>\left( \mid
0_B>+\mid 1_B>\right) \mid 0_C>,\mid \psi _3>=\frac 1{\sqrt{2}}\left( \mid
0_A>+\mid 1_A>\right) \mid 1_B>\mid 0_C>,$

$\mid \psi _4>=\frac 1{2\sqrt{2}}\left( \mid 0_A>-\mid 1_A>\right) \left(
\mid 0_B>-\mid 1_B>\right) \left( \mid 0_C>-\mid 1_C>\right) ,$ in this case 
$\rho _{\left( A,B\right) },\rho _{\left( A,C\right) },\rho _{\left(
B,C\right) },\rho _{\left( A,BC\right) },\rho _{\left( B,CA\right) }$ and $%
\rho _{\left( C,AB\right) }$ all are separable (it can be verified that they
all satisfy the PPT conditions), however $\rho $ is entangled[8,12]. The
source of problem is from the requirement about decomposition of $\rho $ in
the theorem.

(2) The above results in this Letter can be generalized to the more
dimensional cases. Here as an example, we only write the results of
quadripartite qubit density matrixes, the more general extensions are
completely straightforward. Let the natural bases are $\mid i_A>,$ $\mid
j_B>,\mid k_C>$ and $\mid l_D>\left( i,j,k,l=0,1\right) ,$ a quadripartite
qubit density operator

$\rho =\sum\limits_{i,j,k,l,r,s,t,u=0,1}\left[ \rho \right] _{ijkl,rstu}\mid
i_A>$ $\mid j_B>\mid k_C>$ $\mid l_D><r_A\mid <s_B\mid <t_C\mid <u_D\mid ,$
where $\left[ \rho \right] _{ijkl,rstu}$ are the entries of the density
matrix $\rho .$ Now we define 
\begin{equation}
\rho _{\left( A,B\right) }=tr_{CD}\left( \rho \right) ,\rho _{\left(
A,C\right) }=tr_{BD}\left( \rho \right) ,\cdots ,\rho _{\left( CD\right)
}=tr_{AB}\left( \rho \right)
\end{equation}
Next, $tr_D\left( \rho \right) $ is a tripartite qubit density matrix, by
using the symbols in the above case, we can define 
\begin{equation}
\rho _{\left( A,BC\right) }=\left[ tr_D\left( \rho \right) \right] _{\left(
A,BC\right) },\;\rho _{\left( B,CA\right) }=\left[ tr_D\left( \rho \right)
\right] _{\left( B,CA\right) },\;\rho _{\left( C,AB\right) }=\left[
tr_D\left( \rho \right) \right] _{\left( C,AB\right) }
\end{equation}
Similarly, $\rho _{\left( A,BD\right) },\rho _{\left( B,DA\right) },\rho
_{\left( D,AB\right) },\rho _{\left( A,CD\right) },\rho _{\left( C,DA\right)
},\rho _{\left( D,AC\right) },\rho _{\left( B,CD\right) },\rho _{\left(
C,DB\right) },\rho _{\left( D,BC\right) }.$ In addition, we define $\rho
_{\left( AB,CD\right) }$ by 
\begin{equation}
\left[ \rho _{\left( AB,CD\right) }\right] _{ij,rs}=\left[ \rho \right]
_{iijj,rrss}+\left[ \rho \right] _{iijj,rrs\left( 1-s\right) }+\left[ \rho
\right] _{i\left( 1-i\right) jj,rrss}++\left[ \rho \right] _{i\left(
1-i\right) j\left( 1-j\right) ,r\left( 1-r\right) s\left( 1-s\right) }
\end{equation}
Similarly, $\rho _{\left( AC,BD\right) },\rho _{\left( AD,BC\right) }$
(notice that there are several repeated, e.g. $\rho _{\left( BD,AC\right)
}=\rho _{\left( AC,BD\right) },$ etc., hence we delete them). By a similar
way we can define $\rho _{\left( A,BCD\right) }.\rho _{\left( B,CDA\right)
},\rho _{\left( C,DAB\right) },\rho _{\left( D,ABC\right) },$ etc. We can
similarly prove that the above bipartite matrixes all are density matrixes,
the total (except repeated) is 25. And we have completely similar theorem
and corollaries, etc.. Similar way can be directly extended to more high
dimensional case. Sum up, the criteria of separability of multipartite qubit
density matrixes are reduced, in some cases, to the criteria for several
bipartite qubit density matrixes, only the numbers of the corresponding
bipartite density matrixes will increase swiftly.

\end{document}